\documentclass[aps,prl,preprint,superscriptaddress,longbibliography]{revtex4-2}
\usepackage{graphicx}
\usepackage{hyperref}
\usepackage{amsmath}
\usepackage{natbib}
\begin{document}

\title{Photon-number dependent afterpulsing in superconducting nanostrip single-photon detectors}

\author{Sebastian M. F. Raupach}
\email[]{sebastian.raupach@ptb.de}

\affiliation{Physikalisch-Technische Bundesanstalt (PTB), Bundesallee 100, 38116 Braunschweig, Germany}
\author{Mariia Sidorova}
\affiliation{DLR Institute of optical systems, Rutherfordstrasse 2, 12489 Berlin, Germany}
\affiliation{Humboldt-Universit\"at zu Berlin, Department of Physics, Newtonstr. 15, 12489 Berlin, Germany}

\author{Alexej D. Semenov}
\affiliation{DLR Institute of optical systems, Rutherfordstrasse 2, 12489 Berlin, Germany}

\date{\today}

\begin{abstract}
Superconducting nanostrip single-photon detectors (SNSPD) are wide-spread tools in photonic quantum technologies. Here, we study the afterpulsing phenomenon in commercial SNSPD exhibiting significant levels of afterpulses even at low counting rates. We find different contributions, where  the probability of an afterpulse is not a constant but depends on the mean number of photons per light pulse including mean numbers much less than one. Our observations exclude the electrical circuit as the primary cause of the main contribution to the observed afterpulsing probability, which exhibits a strong dependence on mean photon number. We propose a phenomenological model which qualitatively explains our findings via the introduction of slowly relaxing "afterpulsing centers", storing the absorbed photons’ energy for several tens of nanoseconds. We conjecture that two-level systems in amorphous materials are physical candidates for the role of such afterpulsing centers.
\end{abstract}

\maketitle

\section{Introduction}
Single-photon detectors \cite{eisaman2011} are pivotal components of photonic quantum technologies and applications such as quantum key distribution \cite{liao2017, pirandola2020}, photonic quantum computing \cite{arrazola2021, madsen2022}, biophotonics \cite{niwa2017, bruschini2019}, deep-space communication and astronomy \cite{khatri2015, docomas2016, smith2019, deng2019, appel2022} as well as particle physics \cite{bastidon2015} and are likely to become an important tool in time and frequency transfer and metrology \cite{spiess2022}. While superconducting transition edge sensors play an important role in astronomy \cite{smith2019, appel2022}, particle physics \cite{bastidon2015} and photonic quantum computing \cite{madsen2022}, arguably the most wide-spread types of single-photon detector technologies are semiconductor-based avalanche diodes (SPAD) and superconducting nanostrips as single-photon detectors (SNSPD). While SPAD allow for room temperature operation and straightforward integration using standard processes of semiconductor technology, SNSPD require cryogenic operating temperatures of around 3~K or less. SNSPD, on the other hand, exhibit higher detection efficiencies, in particular in the telecom wavelength range and when embedded into an optical stack, as well as a fast response to incoming light pulses and short reset times of nanoseconds, routinely allowing for small timing jitter of the order of 10~ps and count rates of a few 100~MHz \cite{goltsman2001, hadfield2012, sidorova2018, semenov2021, chang2021}.\\
Semiconductor-based single-photon detectors suffer from parasitic trapping and release of charge carriers following a detection event, leading to a temporary increase of dark counts, i.e. detection events in the absence of light but in the wake of another detection event. This behaviour is termed afterpulsing. Afterpulsing in SPAD is typically handled by implementing a hold-off time of the order of microseconds in the detector's driving electronics, to allow for a decay of the parasitic release of charge carriers before the detector is armed again.\\
Afterpulsing in SNSPD is a subtle phenomenon that is usually assumed to be absent or simply ignored but is occasionally observed, e.g., in \cite{fujiwara2011, marsili2012, burenkov2013, kerman2013, miki2017}. There, on a timescale of a few nanoseconds in parallel nanowires, it was attributed to a trade-off between the recovery of the bias current and the instantaneous critical current \cite{marsili2012}, or in single nanowires to technical sources, such as perturbations of the bias current due to reflections in the detector's readout electronics \cite{fujiwara2011,burenkov2013}, or to an overshoot in bias current caused by a discharge current from capacitors in ac-coupled amplifiers after a detection event at high event rates \cite{kerman2013, miki2017}.\\
Considering afterpulsing as a kind of excess noise, one would look for known sources destroying superconductivity locally or introducing electrical noise in superconducting devices. Two-level systems (TLS) are rather ubiquitous objects residing in dielectric layers next to the superconductor such as substrates, intermediate layers as those from an optical stack or oxide layers. In superconducting qubits and superconducting resonators, two-level systems are regarded as being the main factors in decoherence and losses, because of mutual interaction phenomena and dissipative phonon interaction \cite{mueller2019}. Possible effects of two-level systems on the performance of SNSPDs do not yet seem to have found attention.\\
In this article, we investigate afterpulsing behavior in an off-the-shelf commercial SNSPD. We put a particular emphasis on the validity of common explanations for afterpulsing and the possible role of two-level systems. We find, that, while reflections and back-actions of the readout amplifier on the nanostrip may shape the probability distribution of afterpulses over time, they are not the primary cause of the observed afterpulsing phenomenon. In view of the timescales involved and of a significant dependence on the mean photon number per incoming light pulse, we find strong indications that some kind of two- or multi-level system may play a dominant role in the afterpulsing behavior of SNSPD via storage and delayed release of the absorbed photons’ energy. However, further work is required to elucidate the nature of these “afterpulsing centres” and their relation to previously described TLS.

\section{Experimental setup}
\begin{figure}
\includegraphics[width=8.5cm]{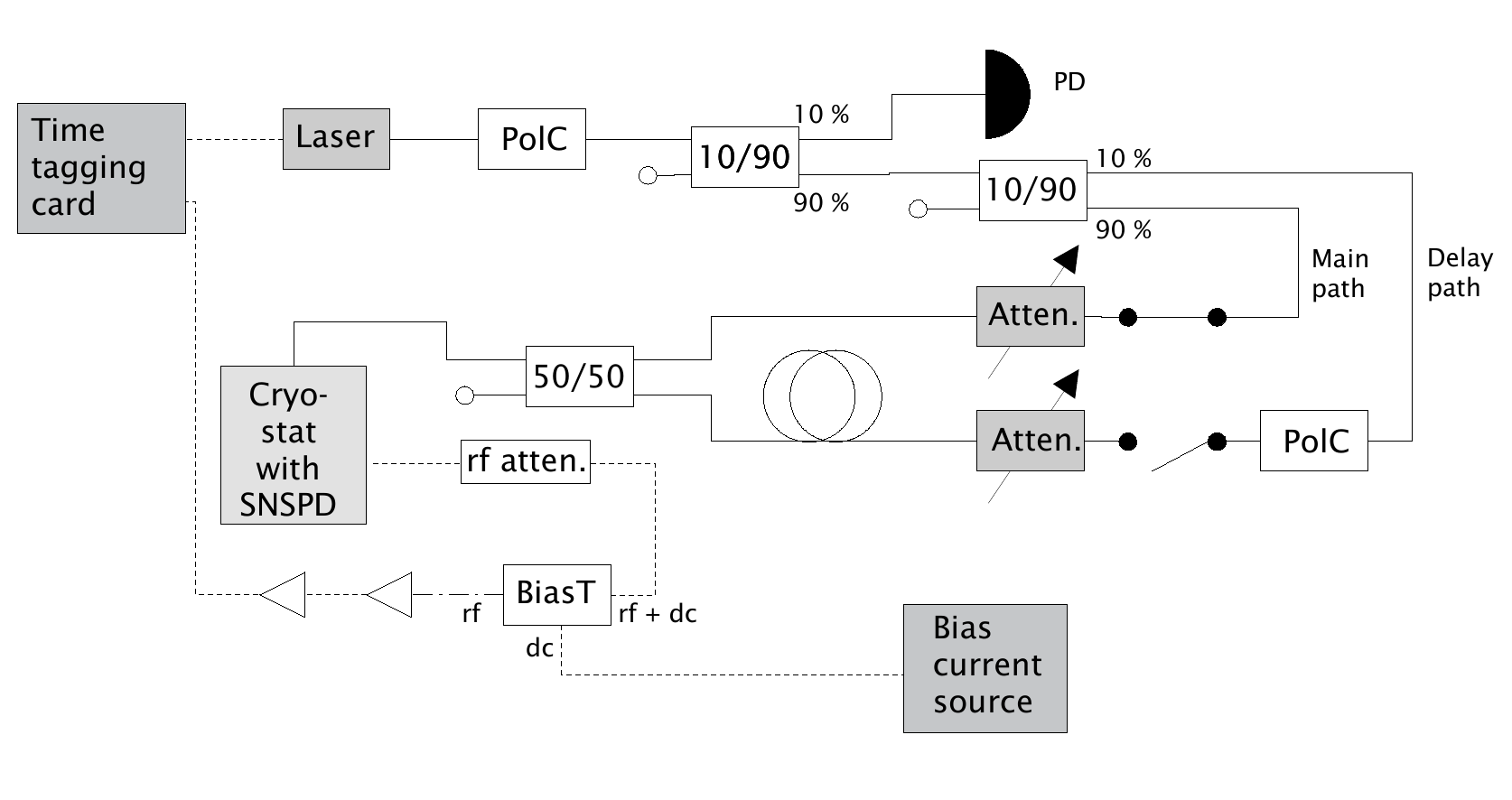}
\caption{Schematic sketch of the experimental setup. The optical setup is fibre-based, where the laser emits light pulses with a duration of less than 100 ps; PolC: manual polarisation controller; Atten.: variable optical attenuator; PD: calibrated photodiode (dash-dotted line: coaxial cables with variable total length).}
\label{fig:setup}
\end{figure}

A schematic sketch of our experimental setup is shown in fig.~\ref{fig:setup}. The fibre-coupled calibrated optical setup follows the ‘double attenuator’ approach \cite{lopez2015}, here comprising a picosecond pulse laser, a calibrated monitor photodiode, and calibrated variable attenuators to provide light pulses with a duration of less than 100 ps at a nominal wavelength of 1548 nm and with known mean numbers of photons per pulse at the detector. The repetition rate of laser pulses is set by a waveform generator (not shown). The setup (and data taking/analysis) is very similar to the one described in \cite{raupach2022}, but here the setup includes manual polarization controllers to maximize the count rate prior to each measurement run due to the SNSPDs’ polarization sensitivity. To also allow for a second, delayed light pulse at SNSPD, we add a second light path (delay path). Each path can be individually attenuated, and opened or blocked using built-in shutters (switch symbols in fig.~\ref{fig:setup}). 
Unless stated otherwise, for the experiments described here only the 'main' path is used. The SNSPD devices are mounted inside a dry vacuum cryostat system (Entropy GmbH) and kept at an ambient temperature of around 2.9~K.

The SNSPD used in this work are two nominally identical, commercial devices (QuantumOpus, 2021) made of an amorphous, unknown material \cite{cahall2017} and undisclosed geometry, where the material is likely to be a silicide or highly disordered nitride. As the manufacturer shared very limited or no technical information regarding these devices’ layout and material which potentially might be relevant, these commercial devices essentially have to be treated as being black-box-like. The devices’ pronounced polarisation sensitivity indicates a meander design. We measured their critical temperature to be around 5.7~K, which is a typical value for amorphous materials. Measurement of the trigger-to-detection jitter of around 500~ps at a bias current of 11.24 $\mu$A (SDE around 20\%) was largely limited by the time resolution of our time tagger (250~ps), which therefore constitutes an upper bound.

One of the two SNSPD (`SNSPD 2') is read out using a cryogenic electronics board provided by the manufacturer and mounted on the 2.9~K stage of the cryostat; the other SNSPD (`SNSPD 1') is read out using commercial room-temperature electronics, consisting of a bias-tee (Mini-Circuits ZFBT-282-1.5A+, 10 MHz to 2.8 GHz) and two amplifiers in series (Mini-Circuits ZX60-3018G-S+, 20 MHz to 3.0 GHz), as well as a 3~dB rf attenuator between bias-tee and cryostat. The purpose of this attenuator is to serve as a well impedance matched room-temperature shunt resistor (Mini-Circuits VAT-3+, 50 $\Omega$ impedance), which we measured to have a throughput  dc  resistance of around 150 $\Omega$ \cite{footnote:shuntresistor}. With  this setup (through the attenuator), and using a sourcemeter (Keithley 2450), we measured an apparent critical (switching) current of SNSPD 1 of at least 13.8~$\mu$A, while the intrinsic switching current of the device actually seems to be smaller \cite{footnote:criticalcurrent}. For the cryogenic readout the observed system detection efficiency (SDE) was around or larger than 85 \% for both devices, for the room temperature readout the SDE over the range of currents used here varied between around 10 \% to around 65 \% (see appendix; for mean photon numbers larger than about 1 despite correcting for deadtime and taking Poisson statistics into account, we saw a systematic decrease in SDE, an observation reported also in \cite{marsili2011} for detectors using a parallel-circuit layout). After verifying that both SNSPDs show comparable afterpulsing behaviour including the dependence on mean photon number, see e.g. fig. \ref{fig:timetraces}, in view of the availability of technical information on the readout electronics required for modeling, subsequent analysis is done solely for results obtained with SNSPD 1  (i.e. using room-temperature electronics). 

\section{Data analysis}
With a time tagging card (TimeHarp 260, Picoquant), each trigger pulse to the laser as well as each count event from the readout electronics (electric pulse) is registered and time tagged (nominal resolution 250 ps). Registered count events, hereafter also counts, originate either from photons absorbed by SNSPD (light counts) or from intrinsic fluctuations in SNSPD (dark counts), which are either conditional afterpulses or equilibrium dark counts. Count histograms have a total time axis equal to the period of laser pulses (interval), i.e., 50 $\mu$s for a repetition rate of 20 kHz. The repetition rates were intentionally chosen to be small to minimize afterpulse count 'overspill' from one interval to another. We analyze the count histograms for each measurement run, e.g., for each mean photon number for a given bias current.\\
To visualize and quantify afterpulsing, we filter the histograms such that only those intervals are included that do contain a detection event (light count) within a certain time window following each trigger pulse to the laser. When selecting the window and deriving the histograms, we accounted for the broadening and shift of the time tags distribution for light counts with the current due to jitter \cite{sidorova2017physical}. As our data showed, a photon count may initiate more than one afterpulse. To distinguish between the first and secondary afterpulses, the histograms are further filtered to only include the first afterpulse following a detection event in the same interval.

Finally, we calculate the distribution of the afterpulsing probability over time by dividing the number of afterpulses in each bin by the total number of intervals contributing to the respective histogram. The latter number equals the total number of detected laser pulses (light counts) in each measurement run.

\section{Observations}

\begin{figure}
\includegraphics[width=9cm]{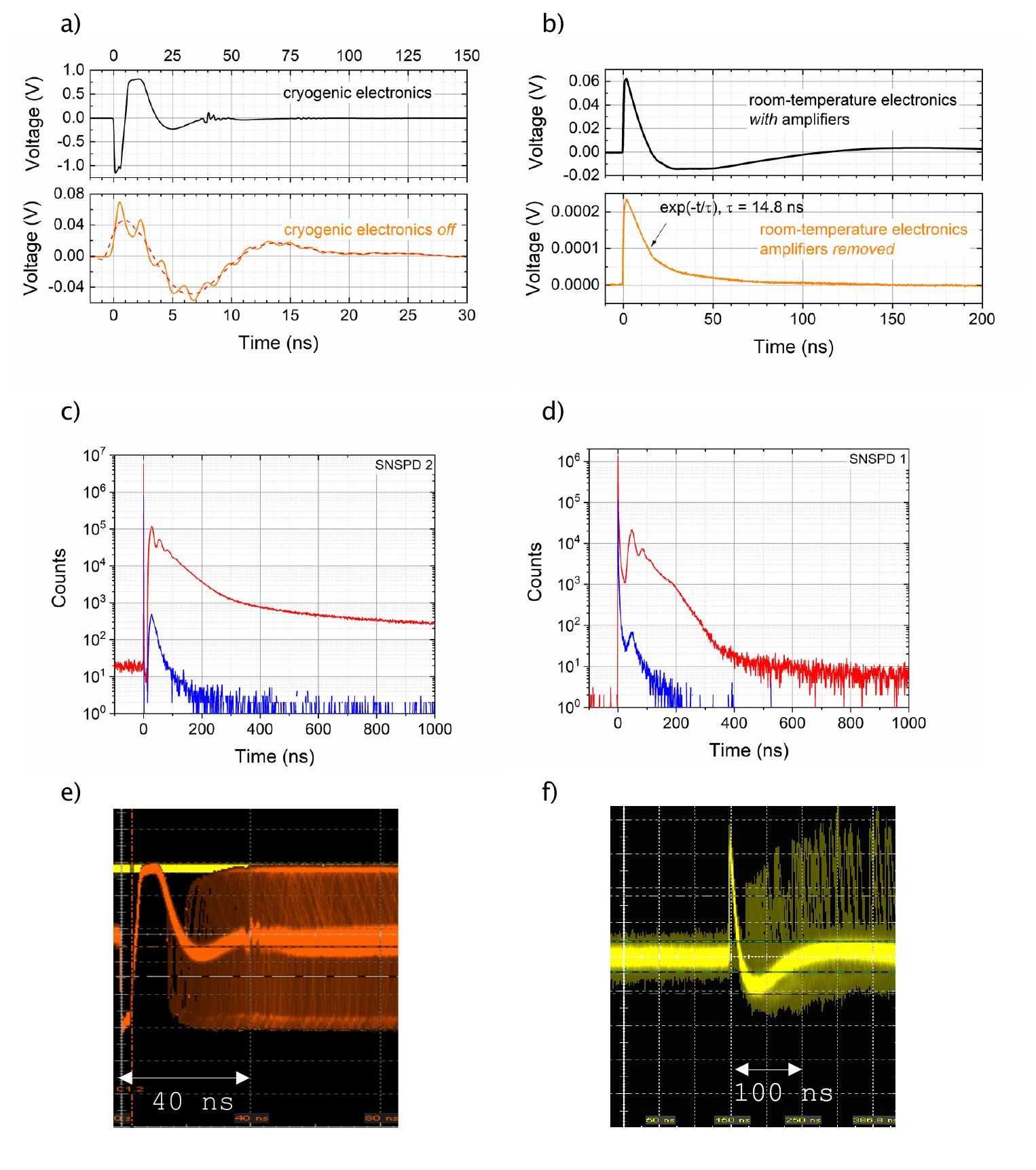}
\caption{Averaged timetrace recorded at the output of (a) the cryogenic electronics of SNSPD 2 (dashed line in the bottom panel: sliding average to guide the eye) and (b) at the output of the room-temperature electronics of SNSPD 1. Count histogram (1~ns bins) of (c) SNSPD 2 (300~s) and (d) SNSPD 1 (150~s) at a bias current of 11.24 $\mu$A, for mean photon numbers per incoming light pulse of 11/pulse (upper red curve) and 0.27/pulse (bottom blue curve). Zero of the time axis is set at the mean onset of light pulses. The repetition rate of light pulses was set to 20~kHz. Panels (e) and (f) show example screenshots of the  oscilloscope in persistent mode showing the light detection pulse followed by afterpulses.}
\label{fig:timetraces}
\end{figure}

Fig. \ref{fig:timetraces} (a) and (b) show (averaged) time traces of the SNSPDs' electrical pulses recorded with an oscilloscope (R\&S RTO2044, nominal bandwidth 6~GHz). The output of the cryogenic electronics board (SNSPD 2, the upper trace in fig.~\ref{fig:timetraces} (a)) shows a sharp voltage peak followed by a broader over- and undershoot voltage. When the power supply to the board is turned off (the lower trace), a train of peaks becomes visible, that are superimposed onto the electrical pulse. The period of this peak train corresponds to the expected round-trip time between SNSPD and the cryogenic electronics board of 1.8~ns, suggesting that there are reflections between them due to some impedance mismatch. This was verified by changing the cable length and observing a corresponding increase in the period. These reflections affect the shape of the recorded traces also when the electronics board is powered.

The output of the room temperature electronics (SNSPD 1, the upper trace in fig.~\ref{fig:timetraces} (b)) shows the typical steep leading edge followed by a slow decay and a voltage under- and overshoot. When removing the amplifiers (the lower trace in fig.~\ref{fig:timetraces} (b)), we do not observe any under- and overshoot, the pulse decays exponentially with a time constant of 14.8~ns.\\
Fig.~\ref{fig:timetraces} (c) and (d) demonstrate that regardless of the type of readout electronics used (with or without visible reflections in time traces), in both cases afterpulsing histograms qualitatively exhibit the same shape, i.e., after the sharp peak corresponding to a light detection event (at zero time), there is an increase in count number due to afterpulsing events followed by a slowly decaying tail. The shape and the height of afterpulsing count histogram as well as the length of its tail change significantly with the mean photon number of the incoming light pulses and with the bias current. The somewhat larger width of the photon peak as well as the faster decay of the afterpulsing tail in the case of SNSPD 1, operated with the 3 dB attenuator as a shunt, likely indicate a different (smaller) current effectively flowing through the SNSPD 1 compared to SNSPD 2, which may be an effect related to the manufacturer’s readout or subtle differences from fabrication. Furthermore, SNSPD2 exhibits a considerably longer afterpulsing ‘tail’ compared to SNSPD1. To further investigate this, for SNSPD 1, we compared the decay of the long-term darkcount level for different mean photon numbers and measurements taken at repetition rates of 20~kHz and 40~kHz to the (constant) count level of ‘dark measurements’ (i.e. shutter closed) routinely taken for each measurement run. Employing a large binning of 1~$\mu$s we found that the level depends significantly on the signal repetition rate (for identical mean photon number) and even for SNSPD 1 remained (very slightly) elevated for signal detection intervals up to about 1 ms. Panels (e) and (f) of fig.~\ref{fig:timetraces} show examples of screenshots of the oscilloscope in persistent mode revealing pulses produced by light counts which are followed by afterpulses, where the early afterpulses exhibit smaller amplitudes.

\begin{figure}
\includegraphics[width=9cm]{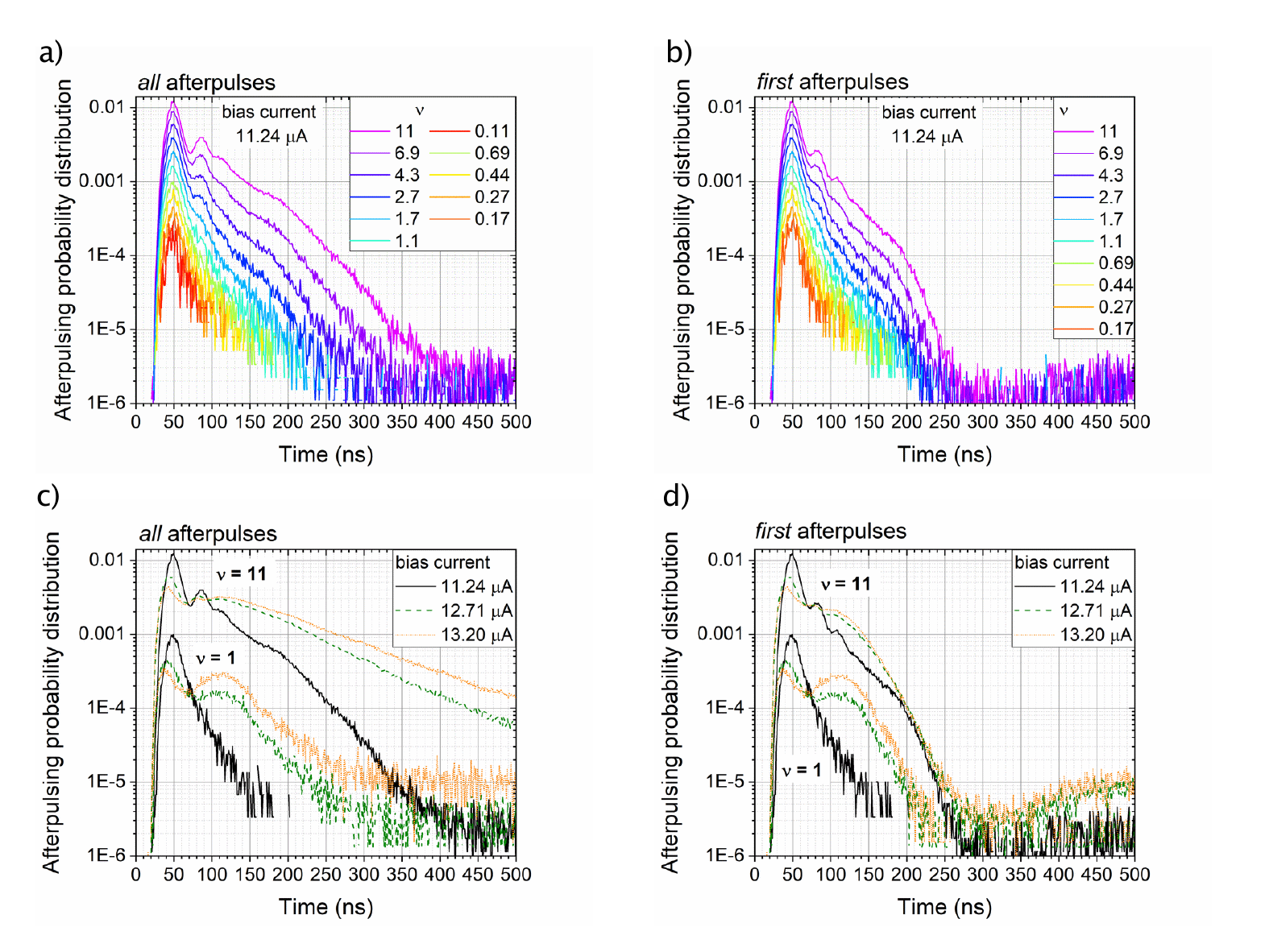}
\caption{Afterpulsing probability distributions for: (a, c) all afterpulses (i.e., including higher-order afterpulses) and (b, d) first afterpulses. The mean number of photons $\nu$ per incoming light pulse as well as the bias currents are indicated.}
\label{fig:afterpulsePD}
\end{figure}

\begin{figure}
\includegraphics[width=9cm]{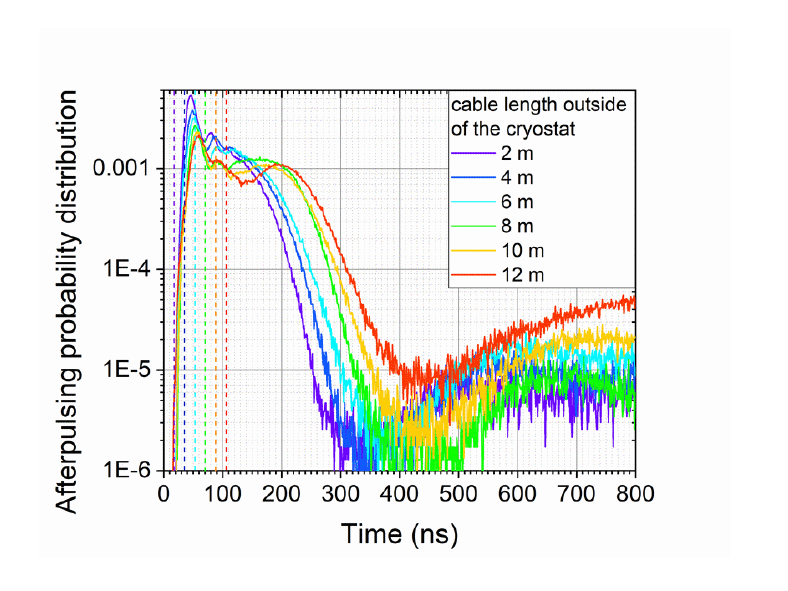}
\caption{Probability distributions of the first afterpulse for different cable lengths between SNSPD and room-temperature amplifiers (see fig. \ref{fig:setup}). The vertical dashed lines indicate a conservative estimate of the round-trip time between SNSPD and amplifiers according to the different cable lengths at room temperature, and based on a measured signal propagation time of 4.4 ns/m.}
\label{fig:cablelength}
\end{figure}

Fig.~\ref{fig:afterpulsePD} displays how the distribution of the afterpulsing probability over time changes with the mean number of photons per incoming light pulse $\nu$ and with the bias current (note that here and in the following only data from SNSPD 1 are shown and discussed). Panels (a) and (c) correspond to distributions for all registered afterpulses (i.e., including higher-order afterpulses) and panels (b) and (d) for only first afterpulses. The width and the height of the afterpulsing probability distribution increases monotonically with mean photon number  $\nu$. For large $\nu$, we find a sequence of arguably three peaks in the distribution of all afterpulses as well as in the distribution of only first afterpulses. These peaks are evidently not due to higher order afterpulsing. The peaks are followed by a decay, which, as expected, is slower when higher-order afterpulses are included. For small $\nu < 1$, the distributions exhibit only one main peak and are the same for both all afterpulses and only first afterpulses.\\
As seen in fig.~\ref{fig:afterpulsePD}(c) and (d), at different bias currents, the dependence of the afterpulsing probability distribution on $\nu$ remains, i.e., the height and the width of the distribution increase with $\nu$. When comparing measurements taken at different signal repetition rates of 20~kHz and 40~kHz (not shown here), we find the amplitude of the first afterpulsing peak to be independent of the repetition rate. However, one can see that currents affect the shape of the distributions which might be the effect of electronics or detector layout \cite{footnote:parallel-layout}.\\
In order to address the effect of electronics, we evaluated the probability distribution of the first afterpulse for fixed $\nu$ and bias current and for different cable lengths between the SNSPD and the room-temperature amplifier. We increased the cable length such that the round-trip time of any electrical pulse would be larger than 110 ns which is far beyond the onset of afterpulsing and the maximum in its probability distribution. The signal propagation time per unit cable length of 4.4 ns/m was measured separately. The results are shown in fig.~\ref{fig:cablelength}. 

In fig.~\ref{fig:cablelength}, vertical dashed lines mark the arrival times of first reflected electrical pulses at the detector for each particular cable length (color encoded). The first falling edge in the probability distribution at around 50 ns delay is not affected by the cable length. Contrary, the second falling edge at around 200~ns moves to larger delay times as the cable length increases. However, the relative delay between this second falling edge and the arrival time of the first reflected pulse does not depend on the cable length, i.e. on the round-trip time. The second falling edge is followed by a by the period of the long-term tail of the afterpulsing distribution, which can be seen to be affected by the cable length as well. \\
At this point, we conclude that although reflections under certain conditions, e.g., for longer cable lengths, may alter the shape of the electrical pulse and also of the afterpulsing probability distribution, they are not the primary cause of the observed afterpulsing.\\
The other common explanation given in the literature for afterpulsing in a single SNSPD is discharge currents from ac-coupled amplifiers. These lead to a temporary increase in the current through the device that overshoots the bias current (‘device current’ $I_d$ in \cite{kerman2013}, fig 1) and result in a corresponding temporary increase in the dark count rate. Since the amplitude of the electrical pulses produced by light and dark counts depends on the current through the SNSPD, in the presence of current overshoots one may expect an increase in the amplitude of afterpulses relative to that of the light detection pulses. In contrast to that, fig.~\ref{fig:timetraces} (f) shows that the amplitude of afterpulses is smaller and then recovers monotonically which strongly questions overshoot currents as being the primary cause of afterpulsing. Actually, the model of \cite{kerman2013} assumes high event rates of at least 10~\% of the inverse recovery time of the detector after a detection event. In our case, where the recovery time is less than 100~ns and the laser repetition rate is 20 kHz, we are several orders of magnitude below these rates.

\begin{figure}
\includegraphics[width=8.6cm]{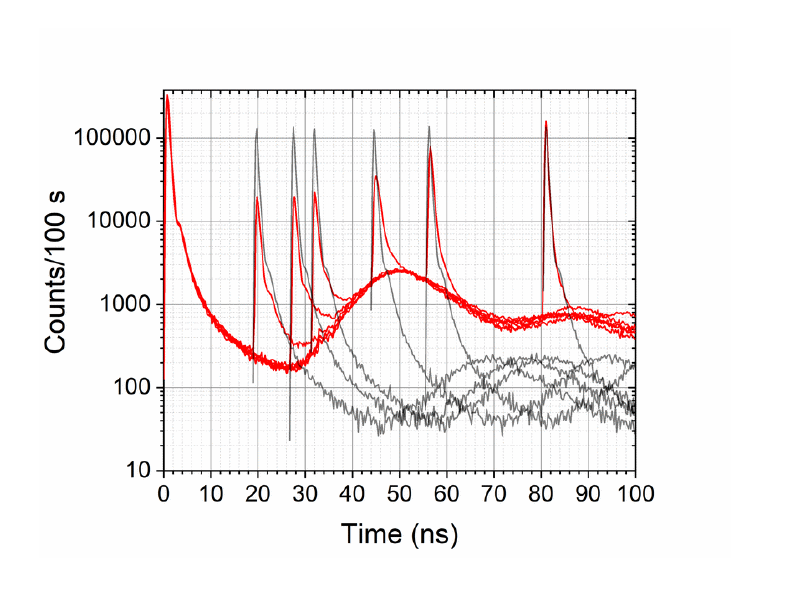}
\caption{Count histograms (here: 250 ps bins) of several `pump-and-probe' type measurements, where the initial light pulse is followed by a second delayed light pulse, for a bias current of 11.24 $\mu$A. Light grey curves: main path blocked, delay path open; red curves: both paths open. The optical delay $t_{d}$ is realized by using in the delay path optical fibres of different lengths $l_{d}$ and their combinations ($l_{d,0}$, $l_{d,0}$+1.5 m, $l_{d,0}$+2.5 m, $l_{d,0}$+ 5.0 m, $l_{d,0}$ + 7.5 m, $l_{d,0}$ + 12.5 m).}
\label{fig:opticaldelays}
\end{figure}

To further address potential overshoot currents, we also perform a 'pump-and-probe' experiment, where the initial light pulse (main path in fig.~\ref{fig:setup}) is followed by a second one (delay path in fig.~\ref{fig:setup}) at a variable delay. The results are shown in fig.~\ref{fig:opticaldelays}. Any overshoot in the otherwise exponential current return to the detector is expected to cause a non-monotonic dependence of the relative detection probability of the second light pulse as a function of the delay time. These relative probabilities correspond to the height of the secondary histogram peaks shown as red curves (both optical paths are open) relative to the black curves (the main path is blocked, while the delay path is open). In contrast to expectations, we see a monotonic increase (recovery) in the detection probability of the second light pulse with the delay time. This holds for all bias currents used here. We should emphasize also that for this 'pump-and-probe' experiment, the round-trip time of an electrical pulse between the amplifier and SNSPD was less than the onset time of afterpulsing (approximately 20~ns). We, therefore, exclude an effect of overshoot currents as well as of the first reflection on the main afterpulsing probability peak and the detection efficiency as a primary cause of afterpulsing. However, given the sparse sampling of delays in the pump-and-probe experiment (fig. \ref{fig:opticaldelays}) one cannot exclude "ringing" in the return current overlaid on its exponential return, even though figure \ref{fig:timetraces} does not indicate it.

\section{Discussion}
We have shown in the previous section, that back action of the readout circuit is not the primary cause of the observed afterpulsing peak. Furthermore, it is distinct from the slow decay of the darkcount level in that it does not change when increasing the repetition rate.\\
Hence, this ‘primary’ afterpulsing contributions appears to be a phenomenon distinct from ‘simple’ fluctuations in current (or temperature), and essentially witness the ability to store and after tens of nanoseconds release the energy of photons being absorbed. Therefore the question remains open, as to this primary cause in view of its dependence on the mean photon number per incoming light pulse.

\begin{figure}
\includegraphics[width=9cm]{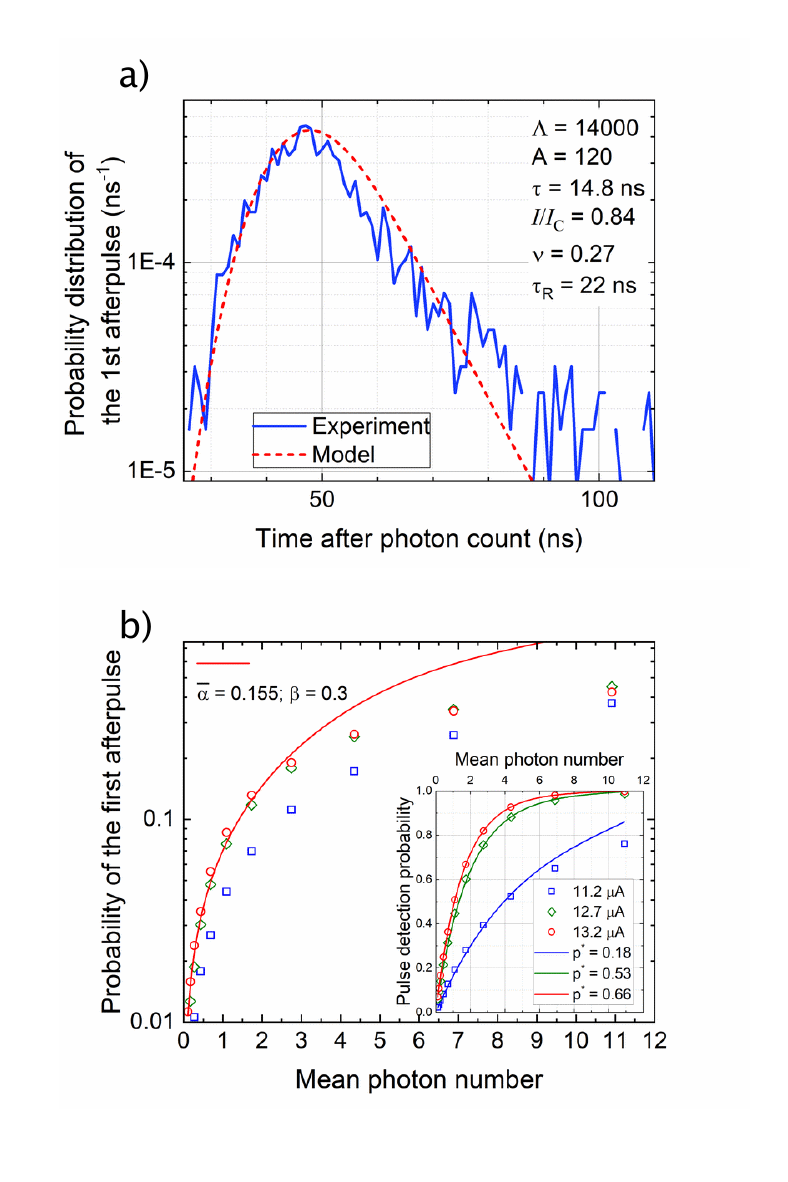}
\caption{(a) Probability distribution of the first afterpulse for $\nu = 0.3$ and the model fit (dashed line). (b) Probability of the first afterpulse as a function of the mean photon number per pulse (symbols) for bias currents 11.2 $\mu$A (square), 12.7 $\mu$A (diamant) and 13.2 $\mu$A (circle). The solid lines shows a qualitative approximation for the bias current 13.2 $\mu$A assuming a $\bar{\alpha}(\nu)$ dependence in eq. \ref{eq:prob_aft}. Inset: Pulse detection probability as a function of mean photon number for the same three currents (legend) and the best fits with $PDP(\nu)$ (see Eq.\ref{eq:prob_aft}), applicable to small $\nu$, where the detector is in the single-photon detection regime. Best fit values of the probabilities $p^*$ to detect a single photon are indicated in the legend.}
\label{fig:probability-first-afterpulse}
\end{figure}

In constructing the microscopic model to follow further below, we concentrate on low mean photon numbers to minimize potential non-linearities due to multi-photon events as well as a potential crosstalk from the afterpulsing’s distribution long tail.\\
Here, we first note that the probability of the first afterpulse after a light count, which is the integral (sum) over time of a corresponding probability distribution (Fig.~\ref{fig:afterpulsePD}b) increases almost proportionally to the mean number of photons $\nu$ in the whole examined range of mean photon numbers. (fig.~\ref{fig:probability-first-afterpulse}(b) and fig.~\ref{fig:probability-first-afterpulse-blackbox-activation}(a)). Given the Poisson distribution of the probability $P_n(\nu)=e^{-\nu}\nu^n/n!$ that exactly $n$ photons arrive at the detector within a random pulse, for the pulse sequence with $\nu \ll 1$ the majority of light pulses arriving at the detector will contain either one or no photons. Furthermore, pulses with one photon will dominate non-empty pulses. For $\nu=0.1$ (lower edge of our experimental range), corresponding relative number are $P_{n\leq1}(0.1)=99$\% and $P_1(0.1)/P_{n\geq1}(0.1)=90$\%. Assuming that for $\nu \ll 1$ afterpulsing as well as pulse detection are both linear responses, i.e. the probability of an afterpulse after a light count and the probability of pulse detection are constant, we come to the conclusion that the afterpulsing probability normalized to the number of detected pulses, $P_A(\nu)$, should saturate at a constant value. This definitely contradicts our experimental observation of the linear dependence of the normalized probability on $\nu$ at small mean photon numbers.

The reason for a $P_A(\nu)\propto\nu$ dependence at $\nu \ll 1$ could be either a significant non-zero correlation between subsequent light counts or non-linearity of the afterpulsing response, i.e. the dependence of the probability of the first afterpulse on the number of photons per pulse. For our typical experimental conditions, the mean time interval between two subsequent non-empty pulses $(\nu F)^{-1}$ ($F$ is the pulse repetition rate) is a few hundreds of microseconds. Although correlation spanning such time interval is feasible, e.g. via heating of the detector holder, it would cause a dependence of the afterpulsing probability on the pulse repetition rate which we did not observe. Invoking the latter nonlinearity, one has to adopt the discreteness of the afterpulsing response. This means that each absorbed photon may initiate an afterpulse independently on others. In order to implement a first approach for the explanation of our experimental data we extend the electro-thermal SNSPD detection model \cite{Semenov2007, Semenov2009, Kerman2009, yang2007} for the multiphoton case and introduce a phenomenological model of afterpulsing centers.

The extended electro-thermal model relies on the assumption that each detected photon initiates a normal domain in the superconducting strip. Although these domains are thermally independent, their dynamics is correlated via the common current through the strip, which is defined by the readout circuit. Mathematical details and the major results are presented in the Appendix~\ref{app:ND_model}. The extended model showed that the lifetime of domains is one order of magnitude less than the characteristic time of the exponential current return in the strip after a count event. Hence, although afterpulses can be initiated by each domain independently, afterpulsing is a retarded response requiring intermediate energy storage.

For the phenomenological (‘microscopic’) model of afterpulsing, we assume that each normal domain activates with the probability $\alpha$ an "afterpulsing center" e.g. in the strip or in the substrate underneath. Each excited center stores a part of the photons’ energy released in the form of a burst of thermal phonons. The centers decay into the non-excited, ground states with a relaxation time $\tau_R$. When returning to its ground state, the center releases the stored energy and may initiate a dark count with the probability $\beta$. On this level, the shape of the probability distribution can be qualitatively reproduced with our phenomenological model and estimated microscopic parameters, where we note that the dependence of the dark count rate in our SNSPD on the relative current $i = I/I_C$ is perfectly described in the framework of the modified theory of thermally activated phase slips \cite{Semenov2020} as $\gamma(i) = \Lambda\, exp(-A (1-i)^{5/4})$ with the critical current $I_C = 13.8\, \mu$A, attempt rate $\Lambda = 14000$ ns$^{-1}$, and the activation exponent $A = 120$. Invoking exponential return of the current in the strip $i(t) = i_0(1 - exp(-t/\tau))$ with the characteristic time $\tau = 14.8$ ns, one can express the probability distribution, $PD(t)$, of the first afterpulse as $PD(t) \propto \gamma(i)exp(-\gamma(i)t)exp(-t/\tau_R)$. In Fig. \ref{fig:probability-first-afterpulse} (a), the dashed curve computed according to this expression with a relaxation time of $\tau_R = 22$ ns represents the best fit to the experimental probability distribution of the first afterpulse.

The probability, $P_m(p^*)$, that exactly $m\leq n$ photons from $n$ arrived at the detector are detected follows the binomial distribution $P_m(p^*) = n!/[(n-m)! m!] (p^*)^m (1-p^*)^{n-m}$ where $p^*$ is the probability for a single photon to be detected. The probabilitiy, $P_k(\alpha)$, that exactly $k$ centers are excited and the probability, $P_j(\beta)$, that exactly $j$ centers initiated an afterpulse also follow binomial distributions. The probability for at least one afterpulse to appear, $P_A(\nu)$ is the sum over all $n, m, k, j\geq 1$ of the product of the Poisson $P_n(\nu)$ and all three binomial distributions. As predicted by the simple physical treatment above, for $\nu\ll 1$, the linear approximation \cite{linear-model-explanation} contradicts experimental data. Mathematical details of the linear approximation and a 'black-box model' fitting all $\nu$ (under the assumption of an activation threshold of one photon) are presented in the appendix~\ref{app: BlackBox_model}.

Elaborating nonlinear approximation requires the dependence of the excitation probability $\alpha$ on $m$. For an arbitrarily dependence, the analytic expression in the closed form is not available. We therefore approximated the sum over all indices by introducing the mean value, $\bar{k}=np^*\bar{\alpha}$, for summations over $m$ and $k$. With this approximation, the sum over $j$ reduces to $1-(1-\beta)^{\bar{k}}$ and the normalized probability of a first afterpulse becomes 

\begin{multline}
P_A(\nu)\approx \dfrac{\Sigma_{n\geq 1}P_n(\nu)(1-(1-\beta)^{np^*\bar{\alpha}})}{\Sigma_{n\geq 1}P_n(\nu)(1-(1-p^*)^n)} \\
= \frac{1-exp\{\nu [(1-\beta)^{\bar{\alpha} p^*}-1]\}}{1-exp\{-\nu p^*\}},
\label{eq:prob_aft}
\end{multline}
where $\bar{\alpha}$ is the weighted average of the excitation probability $\alpha$ over the range of detected photons per light pulse. Note that $p^*<p$ where $p$ is the probability that a single photon is absorbed in the detector. If each absorbed photon may excite an afterpulsing center with the same probability $\alpha$, $p$ should be used instead of $p^*$ in the numerator. The denominator represents the pulse detection probability $PDP(\nu)$, i.e. the probability to detect a light pulse from a random, Poisson-distributed sequence of pulses with the mean photon number pro pulse $\nu$. We used this expression to fit at $\nu\ll 1$ the experimental pulse detection probability defined as $N/K$ where $N$ is the number of detected pulses from a random sequence with $K$ pulses. The fit returns the probabilities $p^*$ to detect a single photon as a function of the bias current. Here they are 0.18, 0.53, and 0.66 for bias currents 11.2 $\mu$A, 12.7 $\mu$A, and 13.2 $\mu$A, respectively. The experimental data and fits are shown in the inset to fig.\ref{fig:probability-first-afterpulse} (b).

For small $\nu$, $P_A(\nu)$ from Eq.~\ref{eq:prob_aft} approaches a constant non-zero value not depending on $\nu$. To address experimental observation, we leave the question of whether and how $\alpha$ and also $\beta$ may depend on the number of simultaneously detected photons to the microscopic picture of excited centers. For the model to agree qualitatively with the experimental data at $\nu < 1$, we let $\bar{\alpha}$ increase with the mean photon number as $\nu^{3/4}$. The solid line in Fig. \ref{fig:probability-first-afterpulse}(b) represents the best fit obtained in this way with eq. 1 for$\nu<1$ and the current 13.2 $\mu$A; values $\bar{\alpha}$ and $\beta$ used for the fit are indicated in the legend in fig.~\ref{fig:probability-first-afterpulse}(b).\\
Below we specifically consider known two-level systems as potential physical candidates for our phenomenological afterpulsing centers. TLS's are well known to affect the performance of kinetic inductance detectors and destroy coherence of superconducting qubits. These findings rely on the cumulative effect of many TLS on the macroscopic properties (e.g. resonance frequency) of microscopically large objects (e.g. strip line resonator). Here we tentatively are looking for a local microscopic action of just a single TLS on the superconducting current in the strip. Consider TLS with a typical level separation \cite{mueller2019} corresponding to a resonance frequency in the GHz range. Assuming e.g. $\Omega = 20$ GHz (equivalent temperature 1 K), at our operating temperature of 2.9 K, approximately 40\% of TLS are expected to populate the excited state. We further assume that the energy $h\Omega$ released by a TLS is directly coupled \cite{Arxiv2018} to the superconducting condensate in the effective fluctuation volume $\xi^2d$ \cite{Semenov2020} where $\xi$ is the superconducting coherence length and $d$ is the film thickness. The kinetic energy of the condensate in this volume is $\varepsilon = L_{sq}I^2(\xi/w)^2/2$ where $L_{sq}$ and $w$ are the kinetic inductance of the square and the width of the strip. The expected relative change of the current caused by the energy released by TLS is $dI/I = h\Omega/4\varepsilon$, where $h$ is the Planck constant. Using a typical coherence length of 5~nm for amorphous Si-based superconductors and the kinetic inductance of the detector (Appendix~\ref{app:ND_model}) we obtain an approximately 10\% increase in the current that drives the volume above the critical state.\\While further experimental work as well as a detailed knowledge of the detector’s layout and material is required to further elucidate the afterpulsing centres’ nature,  our estimate shows that a single TLS may in principle be capable of either causing a deterministic dark count or at least noticeably increasing the effectiveness of intrinsic thermal fluctuations, which are responsible for the background dark count rate \cite{Semenov2020}. The estimate of the TLS activation efficiency $\alpha$ is a more subtle problem. According to the electro-thermal model, the temperature in the center of a normal domain increases up to 15 K (Appendix~\ref{app:ND_model}), but the Joule energy dissipated per one normal domain decreases quickly with the increase of $\nu$ and increases with the current. Assuming that this Joule energy activates TLS results in a decrease in $\alpha$ as a function of $\nu$. On the other hand, the energy released by photons absorbed within the thermal healing length is proportional to the number of such photons. However, at this point identifying the afterpulsing centres with known TLS described in the literature to explain the observed storage and release of the absorbed photons’ energy has to remain a conjecture awaiting amendment, correction or confirmation when further measurements and technical information on these and other devices become available.

\section{Conclusion}
We have investigated afterpulsing in commercial SNSPDs which appear to be particularly sensitive to different sources of afterpulsing, possibly due to their specific layout or material. It appears likely that the detectors’ are made of amorphous material such as silicides or highly disordered nitrides. We found that back-action of the electric readout circuit, which is the conventionally accepted cause of afterpulsing, is not sufficient to explain the observed afterpulsing features listed below. We explained these features invoking microscopic centers which capture and retard thermal energy released in the superconducting strip via photon absorption. Specifically, the features are: (i) the afterpulsing probability grows monotonically with the mean number of photons per incoming light pulse in the range of mean numbers $\nu$ from 0.1 to 10 and this growth appears to only weakly depend on the detection efficiency over a wide range of bias currents; (ii) the relaxation time of microscopic centers responsible for afterpulsing is 22 ns as obtained from the best model fit of the afterpulsing probability distribution; (iii) while the SNSPD remains in the single-photon detection regime, excitation of afterpulsing centers is an essentially non-linear process showing the dominance of multi-photon events in the probability of an afterpulse.

As a plausible candidate for the role of afterpulsing centers we bring forward two-level-systems as a conjecture. The complete microscopic description of afterpulsing, however, requires a detailed knowledge of the properties of TLS or their possible alternatives for a given detector layout and material, which remains a challenge for future work.

We finally note that our observations underline the need for metrology in quantum technologies, i.e. an independent characterization of commercial ‘quantum devices’ to avoid introducing unnoticed systematic errors into a measurement, e.g. when determining the efficiency of single photon sources driven with cw excitation or performing assumedly uncorrelated measurements separated by tens or hundreds of nanoseconds or more. At the same time, given a quantitative microscopic understanding and prediction, the combination of detection and afterpulsing probabilities from a metrological viewpoint might open up exciting new avenues in the calibration of single-photon detectors.

\section{Acknowledgments}
The work reported here was funded by the project EMPIR 19NRM06 METISQ, with additional support from the Quantum Valley Lower Saxony (QVLS) and the cluster of excellence QuantumFrontiers. The METISQ project received funding from the EMPIR programme co-financed by the Participating States and from the European Union's Horizon 2020 research and innovation programme.

We thank Alexey Bezryadin for stimulating discussions, Helmuth Hofer for calibration of the reference detector, Stefan Pendsa and Henry Ganz as well as A. Fern\'andez Scarioni for valuable technical support, and Tim Rambo and Aaron Miller for bringing reference \cite{miki2017} to our attention.

Indicating the name of any manufacturer or supplier by PTB or DLR in this scientific journal shall not be taken to be PTB’s or DLR’s endorsement of specific samples of products of the said manufacturer; or recommendation of the said supplier.

\appendix

\section{Normal domains in the electro-thermal model}
\label{app:ND_model}

\begin{figure}
\includegraphics[width=9cm]{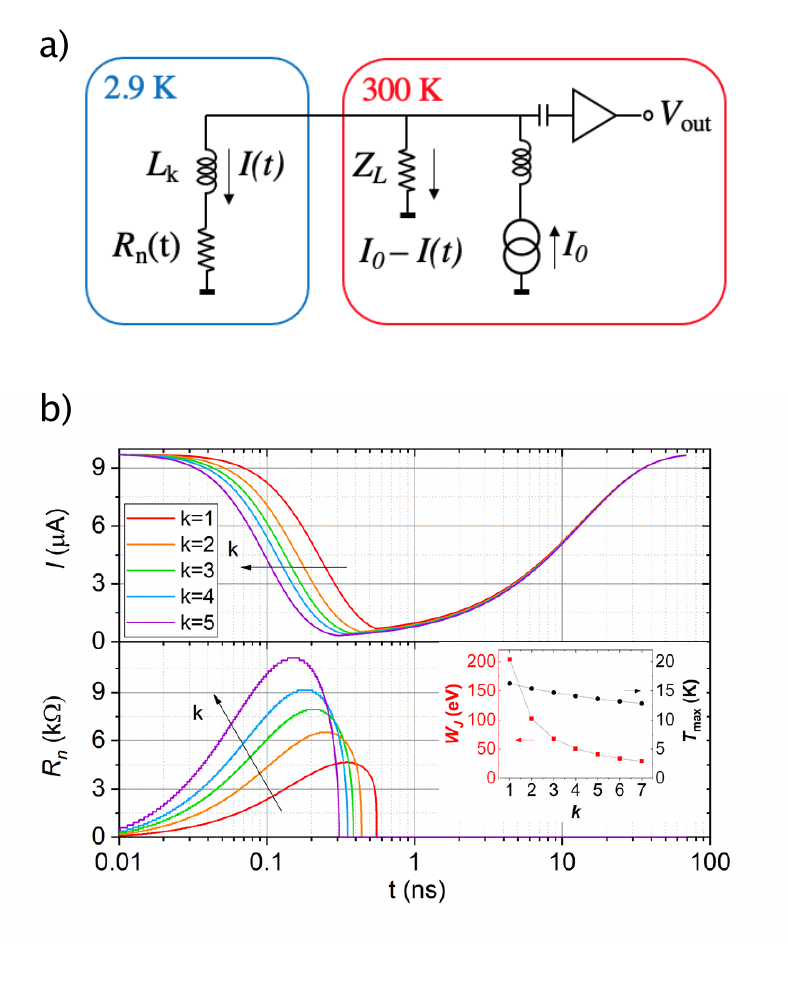}
\caption{(a) Equivalent electrical circuit. (b) Modeled current through the device and total resistance (of all domains) for various numbers of independent domains $k$. Inset: maximum temperature and dissipated Joule emerge per domain vs. the number of domains.}
\label{fig:1a-appendix}
\end{figure}

We derive microscopic inputs for the model from the measured parameters of SNSPD 1, using values of the relaxation time of the electron energy and of the diffusion coefficient typical for superconducting silicides \cite{banerjee2017characterisation} and numerically simulate the evolution of a normal domain and current in a superconducting strip using the electro-thermal model:
\begin{gather}
c_e\dfrac{\partial T}{\partial t} = \dfrac{\partial}{\partial x}(\lambda\dfrac{\partial T}{\partial x}) -K (T^q - T_{bath}^q) + \frac{I(t)^2 R_{sq}}{w^2 d} \nonumber\\
L_{k}\dfrac{dI}{dt} + I(t)[R_n(t) + Z_L] = Z_LI_0,
\label{eq:2TM}
\end{gather}
where the solution of the thermal equation, i.e., the temperature distribution $T(x,t)$ along the strip of the width $w$ and thickness $d$, is coupled to the circuit equation (circuit is shown in fig.\ref{fig:1a-appendix}a) via the time-dependent current $I(t)$ through the strip. In eqns. \ref{eq:2TM}, $T_{bath}$ is the bath temperature, $R_{sq}$ is the square resistance, $c_e$ is the electron heat capacitance, $L_k$ is the total kinetic inductance, $Z_0$ is the load impedance. The thermal equation follows the treatment in \cite{sidorova2022}, with the exponent $q$=3. The electron thermal conductivity $\lambda = D c_e(T_C)$ as well as the effective thermal conductance $K = c_e(T_C)/[q \tau_E(T_C)T_C^{q-1}]$ are taken at the transition temperature, $T_C$, and are the same in the superconducting and in the normal state. Here $D$ is the diffusion coefficient of electrons and $\tau_E$ is the relaxation time of the electron energy to the substrate. Eqns. \ref{eq:2TM} were solved by the finite difference (forward Euler) method as an initial value problem. The size of the seed domain was defined via energy conservation, i.e. a photon with the energy 0.8 eV (1550 nm) heats up the electron and phonon subsystems (with energies defined by the Drude and Debye models) up to temperature $T_C$. For simplicity, the temperature-dependent resistance of the strip was modeled as a step function, $R(T)=0$ below $T_C$ and $R(T)= R_{sq}l/w$ above $T_C$. The transition between the superconducting and normal states was controlled using the Ginzburg-Landau depairing current $I_{dep}(T)$, i.e., if $I \geq I_{dep}(T)$, the state is normal, or otherwise superconducting with zero resistance.

To estimate the relevant parameters of the studied device, we measured critical temperature (5.7 K), device resistance in the normal state (2.3 M$\Omega$), switching current (12.5 $\mu$A, accounting for the rf attenuator), and the electrical signal without amplifier (exponential decay with a time constant of 14.8 ns, the lower trace in Fig.\ref{fig:timetraces} (b)). We assume that the device is a MoSi-based meander with $w=100$ nm, $l=700$ $\mu$m, $d=$ 5 nm, $L_k = 670$ nH, $R_{sq} = 350$ $\Omega$, $D = 0.5$ cm$^2$/s, and $\tau_E = 140$ ps. Here, we assumed that it has a circular layout with a filling factor of 50\% and a diameter of 13 $\mu$m that is typically used for commercial fibre-coupled devices for telecommunication wavelengths.
With these parameters, we numerically solve Eqns. \ref{eq:2TM} for a various number of independent domains created simultaneously (fig. \ref{fig:1a-appendix} (b), semilog-scale). As the domain evolves, the current through the device decreases being diverted into the circuit. When the domain disappears, the current reaches its minimum value and slowly recovers returning back to the strip. The inset to fig. \ref{fig:1a-appendix} (b) shows the maximum temperature in the domain center vs. the number of domains $k$, and the Joule energy dissipated in one domain $W_J = k^{-1} \int I(t)^2 R_n(t) dt$ vs. the number of domains. We estimate that the total dissipated energy elevates the temperature of the device by $\Delta T = W_J k / (c_v V) = 5$~nK, where the phonon heat capacitance per chip volume at 2.9 K is about $c_v \approx 6$ J/Km$^3$ and the chip volume $V \approx$ 1.5 x 10$^-9$ m$^3$. Here we assumed that the substrate material is silicon with a typical thickness of about 300 $\mu$m and that the chip diameter is $\approx$ 2.5 mm (the studied device is coupled to the optical fiber as described in \cite{miller2011}).

From the above analysis, we conclude the following: (i) domain lifetimes do not exceed 1 ns, (ii) the current recovery time does not depend on the number of created domains, (iii) dissipated Joule energy per domain decreases with the number of domains as $W_J \propto 1/k$, (iv) its value is two orders of magnitude larger than the photon energy (1550 nm, 0.8 eV), and (v) it does not lead to any noticeable increase in the chip temperature. 

\section{'Black box' model}
\label{app: BlackBox_model}
\begin{figure}[ht!]
\includegraphics[width=8cm]{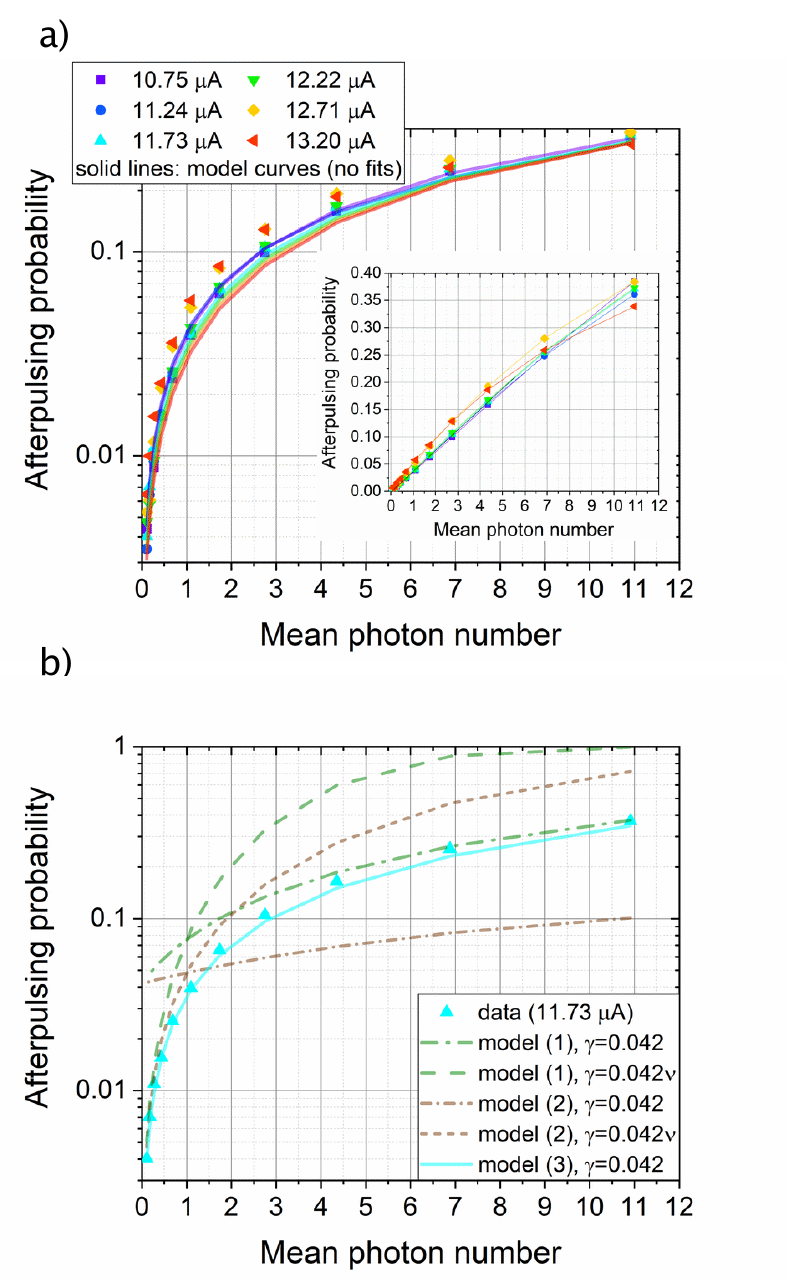}
\caption{Probability of observing at least one afterpulse after a detection event. Panel a) data for a wide range of bias currents and model curves for $\gamma = 0.042$ and a model assuming an 'activation threshold' of one photon (see text) in semi-log presentation; inset: experimental data in linear presentation; panel b) experimental data for a bias current of 11.7 $\mu$A and model curves (see text). }
\label{fig:probability-first-afterpulse-blackbox-activation}
\end{figure}

Here we introduce a phenomenological description of the experimental probability of observing at least one afterpulse treating the detector as a 'black box'. Fig.~\ref{fig:probability-first-afterpulse-blackbox-activation} displays the afterpulsing probability given a detection event over a wide range of bias currents (from 10.75~$\mu$A to 13.20~$\mu$A) (panel a), correponding to detection efficiencies from $\sim 10$~\% to $\sim 65$~\%. The probabilities here were computed by integrating (summation) the corresponding probability distributions, after subtracting the respective distribution's long-term mean value as a background correction \cite{footnote:normalisation}. Over a wide range of currents the probabilities exhibit a very similar and almost linear dependence on the mean number of photons $\nu$ per incoming light pulse, in particular for bias currents of 11.73 $\mu$A and less. A simple linear fit to the data yields a mean slope $\gamma_0$ of around 0.035, which may be interpreted as the linear approximation to an 'afterpulsing efficiency' $\gamma$ for converting a photon into an afterpulse. The actual non-linearity of the probability curve is clearly visible for the largest currents, but is present at all currents. Taking into account the bounding of the afterpulsing probability (first afterpulse) to values $<1$, the probability rather handwaivingly may be approximated by $1-exp(-\gamma\nu)$, where a fit to the data yields a mean value of $\gamma\approx0.042$. While this may serve as a rough back-of-the-envelope estimate of the afterpulsing efficiency, we now set out to systematically develop a phenomenological black box model in detail.

As a first approach we straightforwardly assume that afterpulses are due to the detected fraction $p^*\nu$ of the incoming photons. In general, the probability of an afterpulse is then the probability of a detection event (given a Poissonian input distribution $ \sum_n e^{-\nu}\nu^n/n!$, where the mean photon number is reduced to $p^*\nu$) times the probability that a photon triggers an afterpulse with probability ("efficiency") $\gamma$. For a given photon number $n$ in a pulse, this can be expressed by $1-(1-\gamma)^n$ , where $(1-\gamma)^n$ is the probability that a dark count (afterpulse) is initiated. In this black box model we make no specific assumptions on the microscopic processes inside the SNSPD other than a mutual independence of each photon's effect. Multiplying the probabilities for a given $n$ and summing over all $n$, after proper normalization yields the expression (model 1)
\begin{eqnarray}
P_A(\nu,p^*,\gamma ) &=& \frac{\exp(-p^*\nu)}{1-\exp(-p^*\nu)} \sum_{n=0}^\infty \frac{(p^*\nu)^n}{n!}\left(1-\left(1-\gamma\right)^n\right) \nonumber\\
&=& \frac{1-\exp(-p^*\nu\gamma)}{1-\exp(-p^*\nu)}.
\label{eq:detected_incoming_n}
\end{eqnarray}
The data for a bias current of 11.73 $\mu$A as well as the model curve (brown short-dash-dotted line; $\gamma=0.042$; $p^*(\nu)$ calculated from the pulse detection efficiency) are shown in fig. \ref{fig:probability-first-afterpulse-blackbox-activation}b). We note that also for other values of $\gamma$, this model does not reproduce the observed shape of the probability of observing at least one afterpulse well. If instead assuming that \textit{all} absorbed photons contribute to initiating an afterpulse, we obtain the following expression (here assuming an absorbance of 100 \%; model 2)

\begin{eqnarray}
P_A(\nu,p^*,\gamma ) &=& \frac{\exp(-\nu)}{1-\exp(-p^*\nu)} \sum_{n=0}^\infty \frac{\nu^n}{n!}\left[(1-(1-p^*)^n\right]\cdot \nonumber\\
&& \cdot\left[1-(1-\gamma)^n\right] \nonumber\\
&=& 1 -\exp(-\gamma\nu)\frac{1-\exp(-p^*\nu(1-\gamma))}{1-\exp(-p^*\nu)}.
\label{eq:all_incoming_n}
\end{eqnarray}
The green (upper), dash-dotted line shows the model curve ($\gamma=0.042$). While for large incoming mean photon numbers the model now approximates the data quite well, this is not the case for small mean photon numbers.
Up to now we have assumed the "black-box afterpulsing efficiency" $\gamma$ to be a constant. On the other hand, the energy released by a single absorbed photon might not be sufficient to lead to the initiation of an afterpulse, which might require a "collaborative" behaviour of more than one photon absorptions. As our data are not taken with a single-photon source but with attenuated laser light, this assumption would introduce a dependence of the afterpulsing efficiency on the mean photon number at least for small $\nu$, where photon number states with $n=1$ become dominant over the fraction of all other states where $n>1$ . For illustration, we tentatively replace $\gamma$ by $\gamma\nu$ in equations \ref{eq:detected_incoming_n} and \ref{eq:all_incoming_n}, yielding the dashed model curves in panel b), respectively. As can be seen from the figure, under this assumption both models describe the data considerably better for small mean photon numbers. In particular, the second model, where all absorbed photons contribute to the probability of observing at least one afterpulse, phenomenologically describes the data quite well under the assumption of a linear dependence on the mean photon number for small $\nu$, and a transition region between a photon-number dependent and -independent afterpulsing efficiency at $\nu\approx1$. \\
From these considerations, we now dispense again with the rather arbitrary direct linear dependence on $\nu$, but only assume that all photons absorbed in the SNSPD may contribute to afterpulsing and that the process has an activation threshold corresponding to 1 photon. This yields (model 3):
\begin{eqnarray}
P_A(\nu,p^*,\gamma ) &=& \frac{\exp(-\nu)}{1-\exp(-p^*\nu)} \sum_{n=0}^\infty \frac{\nu^n}{n!}\left[(1-(1-p^*)^n\right]\cdot\nonumber\\
&&\cdot\left[1-(1-\gamma)^ {(n-1)}\right]\nonumber\\
&=& 1 -\frac{\exp(-\gamma\nu)}{1-\gamma}\frac{1-\exp(-p^*\nu(1-\gamma))}{1-\exp(-p^*\nu)}.
\label{eq:all_incoming_n_activation}
\end{eqnarray}
The resulting model curves for $\gamma = 0.042$ are shown in figure \ref{fig:probability-first-afterpulse-blackbox-activation}a). We find that even without fitting $\gamma$, for all $\nu$ model (3) qualitatively describes the experimental data quite well but systematically underestimates the probability of observing at least one afterpulse slightly. This underestimate increases with increasing $p^*$, implying that the overall afterpulsing efficiency $\gamma$ actually increases with the bias current.


\begin{thebibliography}{99}
\bibitem{footnote:shuntresistor} While an inspection of the overall layout of the cryogenic electronics board showed it not to match the information provided by the manufacturer, we were able to identify its shunt resistor (possibly preceded by an inductance) and to measure its dc resistance at cryogenic temperatures (66 $\Omega$).
\bibitem{footnote:criticalcurrent} We also performed a measurement of the switching current by directly connecting the sourcemeter (via a RC low-pass filter of nom. 720 Hz on the current force line) to the detector in a four wire measurement (both with the detector mount grounded to the cryostat and taking care to electrically isolate the detector from the cryostat ground). There we found a critical current (resistance step) at an input current of around 2.5~$\mu$A and thermal runaway setting in either directly (no RC filter) or at around 3.4~$\mu$A (with RC filter), as witnessed by the detector becoming fully normally conducting. Therefore, compared to other SNSPD, the shunt resistance might play a more prominent role as a current divider to set the working point of the device given a certain input current.
\bibitem{linear-model-explanation} Linear means that (i) the mean number of microscopic centers activated by many absorbed photons is proportional to the number of absorbed photons and the proportionality coefficient, $\alpha$, does not depend on the number of photons; (ii) the mean number of initiated afterpulses is proportional to the number of activated centers and the proportionality coefficient, $\beta$, does not depend on the number of activated centers.
\bibitem{footnote:normalisation}We note that for the largest mean photon number and in particular for the largest currents the long afterpulsing ‘tail’ this subtraction may lead to a slight underestimate of the afterpulsing probability.
\end{thebibliography}

\end{document}